\documentclass[prb,showpacs,twocolumn,amsmath,amssymb]{revtex4}
\usepackage{epsfig}
\usepackage{epsf}
\usepackage{amssymb, amsmath}
\usepackage{graphicx}
\usepackage{dcolumn}
\usepackage{bm}
\usepackage{appendix}

\newcommand{\up}{\uparrow}
\newcommand{\dn}{\downarrow}

\newcommand{\bfk}{{\bf k}}

\usepackage{color}
\usepackage{ulem}
\usepackage{amsthm}

\begin{document}

\title{A theorem regarding families of topologically non-trivial fermionic systems}
\author{Bruno Mera$^{1,3}$,  Miguel A. N. Ara\'ujo$^{1,2}$
and V\'{\i}tor R. Vieira$^1$}

\affiliation{$^1$ CeFEMA, Instituto Superior
T\'ecnico, Universidade de Lisboa, Av. Rovisco Pais, 1049-001 Lisboa, Portugal}
\affiliation{$^2$  Departamento de F\'{\i}sica,  Universidade de \'Evora, P-7000-671, \'Evora, Portugal}
 \affiliation{$^3$  Physics of Information Group, Instituto de Telecomunica\c{c}\~oes, 1049-001 Lisbon, Portugal}

\begin{abstract}
We introduce a Hamiltonian for fermions on a lattice and prove a theorem regarding its topological properties. We identify the topological criterion as a $\mathbb{Z}_2-$ topological invariant $p(\bfk)$ (the Pfaffian polynomial). The topological invariant is not only the first Chern number, but also the sign of the Pfaffian polynomial coming from a notion of duality. Such Hamiltonian can describe non-trivial Chern insulators, single band superconductors or multiorbital superconductors. The topological features of these families are completely determined as a consequence of our theorem. Some specific model examples are explicitly worked out, with the computation of different possible topological invariants.
\end{abstract}


\maketitle

\section{ Introduction}
\label{sec: I}

The topological properties of fermionic systems have been the focus
of intensive research since Haldane's seminal work on the
anomalous Hall insulator in the honeycomb lattice
and its extension to time-reversal invariant (TRI) systems\cite{haldane88,kanemele}.
A topological system has a bulk gap characterized by a 
topological invariant. The bulk-boundary
 correspondence  then establishes that edge states exist 
 at the boundary between regions with different topological indices.
 The topological systems of interest include  both
insulators\cite{hasankane} and superconductors\cite{QiZhang,alicea}.

For topological superconductors, the emergence of  zero energy excitations  
which are their own antiparticles, or Majorana fermions (MFs), has long been 
predicted\cite{kitaev}.
Early theoretical models of two-dimensional topological superconductivity
 consider $p+ip$  pairing in an otherwise trivial band.  
Non-abelian MF's  were shown to arise
at vortex cores in a model of spinfull fermions with spin  triplet $p+ip$  pairing 
and where  the two spin components effectively   decouple\cite{ivanov}. 
The pursuit for Majorana fermions as emergent quasi-particle excitations in condensed matter systems 
is very exciting not only from the theoretical point of view but also because it provides a path to realize 
fault-tolerant topological quantum computation\cite{QiZhang,alicea,nayak}.

Physical realizations of topological materials include real materials, such as the
three-dimensional topological superconductor
Cu$_x$Bi$_2$Se$_3$. 
A relevant example of a multiband superconductor
believed to have   $p+ip$ symmetry is Sr$_2$RuO$_4$
\cite{kallin,yada2014}.
The possibility of engineering topological superconductors 
using 
proximity coupling of the
 surface of a three-dimensional topological insulator (TI) or a two-dimensional semiconductor
in proximity to a s-wave superconductor was discussed\cite{engeneering,nature}.
Other possibilities  include time-periodic driving by laser fields\cite{oka,kitagawa,lindner},
optical lattices\cite{optical} and photonic crystals\cite{photonic}.

One important research line  has been the search for band  models 
displaying  non-trivial topology\cite{dasSarma1,mudry,wen,Fito,pequim-1}.
We note that single band models with   $p+ip$  (or other) type of pairing 
 are a theoretical simplification, as topological materials are necessarily multiorbital. 
Some theorems on  the topological indices 
to be expected for various superconductor models
have recently been established.
Symmetries, such as 
lattice inversion symmetry or time-reversal invariance play an important role.
 In many cases, the topology of the Fermi surface (FS) 
is itself  important\cite{eu2015}.
For instance, 
under the assumption of 
lattice inversion symmetry, TRI
and odd parity pairing,
three-dimensional superconductors are  topological if they possess 
an odd number of Fermi surface pockets\cite{Fu1,Fu2}.
A theorem relating the topological indices of a
superconductor to the FS topology
has been established by Sato\cite{sato2010} for the case
where the normal bands have inversion symmetry and pairing
has odd parity.
In the case of time-reversal invariant
single band spin triplet superconductors,
the topological indices were also shown to be related to
FS topology\cite{sato2009}.
Models of two-dimensional superconductors with a pseudospin
degree of freedom have been proposed recently, concentrating on the case
of nodeless odd parity pairing in TRI superconductors. In this case
non-trivial topology requires spin-orbit couplings non-diagonal
in the pseudospin channel\cite{viola1,viola2}.

The paper is structured as follows. In section \ref{sec: II}, we introduce a Hamiltonian for fermions on a lattice 
[see equation \eqref{eq: brunoH} below]
and prove a theorem 
on its topological properties. In section \ref{sec: III}, we discuss several classes of physical systems which can be described by this Hamiltonian, providing some specific model examples illustrating the results. In section \ref{sec: IV}, we provide a summary and conclusions, including a brief discussion regarding the consequences of bulk-boundary correspondence principle in the physical systems described by our model Hamiltonian.

\section{Theorem}
\label{sec: II}

We want to study a class of Hamiltonians, described by smooth families of Hermitian matrices parametrized by some smooth manifold $M$, $\{H(p):p\in M\}$, which are non-singular, i.e., $\det H(p)\neq 0$ for all $p\in M$. This is physically equivalent to a gap condition between occupied and unoccupied bands. Since we are interested in the topological nature of these bands, it is natural to introduce an equivalence relation between Hamiltonians which basically identifies Hamiltonians which can be smoothly deformed into each other without closing the gap, while preserving the notion of occupied and unoccupied bands. Equivalently, two Hamiltonians are equivalent if they can be smoothly deformed into each other without violating the condition $\det H\neq 0$ and preserving the collections of eigenspaces (more precisely, these collections are families of eigenspaces parametrized by the manifold $M$) with positive eigenvalues and negative eigenvalues (with the convention that occupied states have negative energy and unoccupied states have positive energy). One operation which is quite natural, then, is to flatten the spectrum of a given Hamiltonian. In the simplest case, if $\epsilon_1, ...,\epsilon_n$ are the eigenvalues of a given Hermitian $n\times n$ matrix $H$ (the trivial family), we can, if $\det H\neq 0$, smoothly deform the matrix $H$ into a new matrix $\tilde{H}$ which has eigenvalues $\pm 1$ just by taking $\epsilon_i (t)=(1-t)\epsilon_i + t\epsilon_i/|\epsilon_i|,\ i=1,...,n$. It is clear that $H$ and $\tilde{H}$ are equivalent under the equivalence relation described before. This operation will be useful later on.

We consider that our physical system enjoys translation invariance so that momentum $\bfk$ 
is a good quantum number. Then, the class of Hamiltonians we are interested in is described by  smooth families of $4\times 4$ Hermitian matrices parametrized by the set of all possible momenta $\textbf{k}$. The latter is the Brillouin zone which is naturally identified with a two-torus $\mathbb{T}^2$, which is a two-dimensional smooth manifold. In the notation of the beginning of this section, an arbitrary Hamiltonian in our class is described by the smooth family $\{H(\bfk): \bfk \in \mathbb{T}^2\}$. For an arbitrary momentum $\bfk$, the Hamiltonian $H(\bfk)$ is a matrix of the form
\begin{align}
H(\bfk)=
\left(\begin{array}{cc}
h(\bfk)\cdot\tau & i D^{*}(\bfk)\tau_2\\
-iD(\bfk)\tau_2 & -(h(\bfk)\cdot \tau)^T
\end{array}
\right) 
\,,
\label{eq: brunoH}
\end{align} 
where $\tau\equiv(\tau_0,\tau_1,\tau_2,\tau_3)$ is a vector of Pauli matrices with the convention
that $\tau_0$ denotes the two-dimensional identity matrix, $h\equiv(h_0,h_1,h_2,h_3)$ is a vector in four-space and $D=D_1\pm iD_2$ is  complex. 
The notation $h\cdot\tau$ denotes the sum $\sum_{\mu=0}^3 h_\mu\tau_\mu$.
The nature of these matrices is more naturally understood once we introduce the vectors $\vec{h}\equiv (h_1,h_2,h_3)$ and $\vec{h}'\equiv (D_1,D_2,h_0)$, both in three-space. The Hamiltonian can then be written as
\begin{eqnarray}
H(\bfk)=\vec{h}(\bfk)\cdot \vec{S}+\vec{h}'(\bfk)\cdot \vec{T},
\end{eqnarray}
where $\vec{S}\equiv (S_1,S_2,S_3)$ and $\vec{T}\equiv (T_1,T_2,T_3)$ are both arrays of matrices, $\vec{h}\cdot\vec{S}=\sum_{i=1}^3 h_i S_i$ and $\vec{h}'\cdot\vec{T}=\sum_{i=1}^3 h_i ' T_i$. Explicitly,
\begin{align}
S_1=\left(\begin{array}{cc}
\tau_1 & 0\\
0 & -\tau_1
\end{array}\right),\ S_2=\left(\begin{array}{cc}
\tau_2 & 0\\
0 & \tau_2
\end{array}\right),\ S_3=\left(\begin{array}{cc}
\tau_3 & 0\\
0 & -\tau_3
\end{array}\right), \nonumber \\ 
T_1=\left(\begin{array}{cc}
0 & i\tau_2\\
-i\tau_2 & 0
\end{array}\right),\ T_2=\left(\begin{array}{cc}
0 & \tau_2\\
\tau_2 & 0
\end{array}\right),\ T_3=\left(\begin{array}{cc}
I & 0\\
0 & -I
\end{array}
\right).
\end{align}
These matrices are Hermitian generators of a $\mathfrak{su}(2)\oplus\mathfrak{su}(2)$ Lie algebra,
\begin{eqnarray}
[S_i,S_j]=2i\varepsilon_{ijk}S_k,\ [T_i,T_j]=2i\varepsilon_{ijk}T_k,\ [S_i,T_j]=0,
\end{eqnarray}
where $\varepsilon_{ijk}$ is the Levi-Civita totally anti-symmetric symbol (note that the above commutation relations are the same as those of the Pauli matrices $\sigma_1,\sigma_2,\sigma_3$ associated with spin-$\frac{1}{2}$). Additionally, we have Clifford algebra relations,
\begin{eqnarray}
\{S_i,S_j\}=2\delta_{ij}I,\ \{T_i,T_j\}=2\delta_{ij}I,
\end{eqnarray}
where $I$ denotes the $4\times 4$ identity matrix. The Lie algebra $\mathfrak{su}(2)\oplus\mathfrak{su}(2)$ is isomorphic to the Lie algebra $\mathfrak{so}(4)$ of anti-symmetric $4\times 4$ matrices. There is a natural linear transformation of duality in the Lie algebra $\mathfrak{so}(4)$ and this operation identifies self-dual and anti-self-dual generators which then provide the splitting $\mathfrak{so}(4)=\mathfrak{so}(3)\oplus\mathfrak{so}(3)\cong\mathfrak{su}(2)\oplus\mathfrak{su}(2)$. The generators $S$ are self-dual, meaning that the operation of duality acts by multiplication by $+1$, while the generators $T$ are anti-self-dual, meaning that the duality operation acts by multiplication by $-1$. The Lie group $\text{SO}(4)$ acts by conjugation on its Lie algebra and the orbits are completely determined by two $\text{SO}(4)-$invariant polynomials: the Pfaffian of the matrix and its norm. The orbits can be either diffeomorphic to the two-sphere, $S^2$, or to the cartesian product of two two-spheres, $S^2\times S^2$. Via the Hamiltonian equivalence introduced before, there are only two inequivalent classes of orbits. One with positive Pfaffian and the other with negative Pfaffian. One can choose a representative of each of these classes to be a two-sphere $S^2$.\\

The relevant quantity to be studied is the sign of the polynomial $p(\bfk)\equiv |\vec{h}(\bfk)|^2-|\vec{h}'(\bfk)|^2=(|\vec{h}(\bfk)|-|\vec{h}'(\bfk)|)(|\vec{h}(\bfk)|+|\vec{h}'(\bfk)|)$, which is naturally associated with the Pfaffian polynomial in a representation of $H(\bfk)$ by a skew-symmetric matrix in $\mathfrak{so}(4)$ (namely $iH(\bfk)$ can be recast as a skew-symmetric matrix by a similarity transformation). The polynomial $p(\bfk)$ is simply a square root of the determinant of $H(\bfk)$. By the condition $\det H(\bfk)\neq 0$, $p(\bfk)$ has always the same sign and, thus, we can not deform a Hamiltonian with positive $p(\bfk)$ to a Hamiltonian with negative $p(\bfk)$. On the other hand, we can deform a Hamiltonian with arbitrary positive (negative) $p(\bfk)$ so that $p(\bfk)\equiv 1$ ($p(\bfk)\equiv -1$). If $p(\bfk)$ is positive (negative), then, by spectrum flattening we can reduce it to an (anti-)self-dual Hamiltonian. The unitary matrix $U(\bfk)$ which takes $H(\bfk)$ to diagonal form $U^{\dagger}(\bfk)H(\bfk)U(\bfk)=\text{diag}(\epsilon_1(\bfk),...,\epsilon_4(\bfk))$, which can be naturally associated with a rotation $R(\bfk)$ in $\text{SO}(4)$, can be written as,
\begin{eqnarray}
U(\bfk)=\exp(i\vec{x}(\bfk)\cdot\vec{S}/2)\exp(i\vec{y}(\bfk)\cdot \vec{T}/2),
\end{eqnarray}
where $\vec{x}(\bfk)$ and $\vec{y}(\bfk)$ are three-space vectors. Notice that because of the algebraic properties of $\vec{S}$ and $\vec{T}$, we have,
\begin{align}
\exp(i\vec{x}(\bfk)\cdot\vec{S}/2)&= \nonumber \\
&=\cos(|\vec{x}(\bfk)|/2)I+i\sin(|\vec{x}(\bfk)|/2)\frac{\vec{x}(\bfk)\cdot\vec{S}}{|\vec{x}(\bfk)|},\\
\exp(i\vec{y}(\bfk)\cdot\vec{T}/2)&= \nonumber \\
&=\cos(|\vec{y}(\bfk)|/2)I+i\sin(|\vec{y}(\bfk)|/2)\frac{\vec{y}(\bfk)\cdot\vec{T}}{|\vec{y}(\bfk)|}.
\end{align}
The matrix $U(\bfk)$ carries all the data of the eigenvectors of $H(\bfk)$, namely its columns are the eigenvectors themselves, and it is well-defined up to multiplication on the right by gauge transformation of the form,
\begin{eqnarray}
g(\theta,\eta)=\exp(i\theta S_3/2)\exp(i\eta T_3/2).
\end{eqnarray}
If the Hamiltonian is self-dual, then the part of $U(\bfk)$ which is exponential of anti-self-dual generators act trivially. If the Hamiltonian is anti-self-dual, then the part of $U(\bfk)$ corresponding to the exponential of self-dual generators acts trivially. [This is due to the relations $[S_i,T_j]=0$.] Now, if $p(\bfk)$ is positive, we can smoothly deform $H(\bfk)$ to a self-dual Hamiltonian and, similarly, if $p(\bfk)$ is negative, we can smoothly deform $H(\bfk)$ to an anti-self-dual Hamiltonian. By the previous argument, the matrix $U(\bfk)$ which brings $H(\bfk)$ to diagonal form will act independently on self-dual and anti-self-dual generators, rotating the vectors $\vec{h}$ and $\vec{h}'$ independently, so that they are aligned with the $Z$-axis. The same matrix that rotates $\vec{h}$ ($\vec{h}'$ in the anti-self-dual case) to be parallel to the $Z$-axis will rotate the deformed unit vector $\widetilde{\vec{h}}$ ($\widetilde{\vec{h}'}$ in the anti-self-dual case) after spectrum flattening. The reason is because the matrix $U(\bfk)$ is preserved in the spectrum flattening deformation and the fact that $U(\bfk)$ acts independently on $S$ and $T$ generators. Thus we conclude that the resulting deformed Hamiltonian is $\widetilde{H}(\bfk)=\vec{h}(\bfk)\cdot \vec{S} /|\vec{h}(\bfk)|$ in the self-dual case and $\widetilde{H}(\bfk)=\vec{h}'(\bfk)\cdot\vec{T}/|\vec{h}'(\bfk)|$ in the anti-self-dual case. With this construction, the winding number of the induced map to the two-sphere given by the unit vector $\vec{h}(\bfk)/|\vec{h}(\bfk)|$ or $\vec{h}'(\bfk)/|\vec{h}'(\bfk)|$ yields the first Chern number of the Bloch bundle (with ``negative'' energy, i.e., occupied band) of the Hamiltonian. With this, we have proved the theorem which we will now state concisely to finish this section.

\newtheorem*{theorem*}{Theorem}
\begin{theorem*} 
If a physical system is modeled by a smooth family of Hamiltonians $H(\bfk)$, where $\bfk$ denotes momentum in the Brillouin zone, of the form \eqref{eq: brunoH}, such that the condition $\det H(\bfk)\neq 0$ is satisfied for all momenta, then the following statements hold:
\begin{itemize}
\item[(i)] The polynomial $p(\bfk)$ is either positive or negative for all $\bfk$, and $H(\bfk)$ can be smoothly deformed into $\widetilde{H}(\bfk)=\vec{h}(\bfk)\cdot \vec{S}/|\vec{h}(\bfk)|$ or $\widetilde{H}(\bfk)=\vec{h}'(\bfk)\cdot\vec{T}/|\vec{h}'(\bfk)|$, correspondingly;
\item[(ii)] The first Chern number of the relevant Bloch bundle is given by twice the winding number of the map $\Phi_1:\bfk\mapsto \vec{h}(\bfk)/|\vec{h}(\bfk)|$ if $p(\bfk)>0$, or by twice the winding number of the map $\Phi_2:\bfk\mapsto \vec{h}'(\bfk)/|\vec{h}'(\bfk)|$ if $p(\bfk)<0$;
\item[(iii)] In the case where $H(\bfk)$ is a Bogoliubov-de Gennes (BdG) Hamiltonian then one has to account for the doubling of the degrees of freedom (particles and holes), and thus the first Chern number of the relevant Bloch bundle does not have the factor of $2$. This means that the first Chern number of the Bloch bundle is given by the winding number of $\Phi_1$ if $p(\bfk)>0$, or by the winding number of $\Phi_2$ if $p(\bfk)<0$. Note that if $H(\bfk)$ is to describe a BdG Hamiltonian, then all the functions appearing in the matrix $H(\bfk)$ must be even by particle-hole symmetry. 
\end{itemize}
\label{th: 1}
\end{theorem*}

We point out that in the hypothesis of our theorem there is no requirement on inversion symmetry or parity.

The above theorem is  an extension of previous models for practical realization of Majorana modes 
(see, for instance, Section III B of the review by Alicea \cite{alicea}), which did not identify  the topological criterion 
found as a $\mathbb{Z}_2-$ topological invariant $p(\bfk)$ (the Pfaffian polynomial), as we did. 
In the above model (\ref{eq: brunoH})
the topological invariant describing the system is not only the first Chern number, but also the sign of the Pfaffian polynomial coming from this notion of duality we have discussed.
 In other words, one can not go from a system with positive $p(\bfk)$ to a system with negative $p(\bfk)$ without 
 closing the gap.

\section{Applications}
\label{sec: III}

We now discuss the physical systems which can be described by the above 
Hamiltonian form.

\subsection{Topological insulator}

We will first show that the family of Hamiltonians $H(\bfk)$ as in the hypothesis of the theorem
 \textit{cannot} describe a non-trivial topological insulator. 
Consider first that  the matrix  (\ref{eq: brunoH})  is written in the basis
$(\psi_{1\up} \psi_{1\dn}  \psi_{2_\up}\psi_{2\dn})$, for fermions on a lattice
with two orbitals per site (labeled by the subscripts 1 and 2). 
The time reversal  operator  reads
\begin{eqnarray}
\mathcal{T}: H(\bfk)\mapsto (\tau_0\otimes\sigma_y)\cdot H^*(-\bfk)\cdot (\tau_0\otimes\sigma_y),
\end{eqnarray}
where the notation $\sigma_y\equiv \tau_2$ is used to emphasize the physical spin nature of this degree of freedom. 
The invariance under time reversal yields the conditions,
\begin{align}
&\vec{h}(-\bfk)=-\vec{h}(\bfk),\ h_0(-\bfk)=h_0(\bfk) \nonumber \\
&\text{and } D(-\bfk)=D^{*}(\bfk).
\label{eq: TRS constraints (i)}
\end{align}
The condition on $\vec{h}(\bfk)$ implies that, at time reversal invariant (TRI) points, $\vec{h}(\bfk)$ is identically zero. Since the sign of $p(\bfk)$ is the same at each point of the Brillouin zone, the family of Hamiltonians must have $p(\bfk)<0$, otherwise the gap condition would be violated at these points. 
As such, by point $(i)$ of the Theorem,
 the Hamiltonian can be deformed into an anti-self-dual Hamiltonian of the 
 form $\widetilde{H}(\bfk)=\vec{h}'(\bfk)\cdot \vec{T}/|\vec{h}'(\bfk)|$. 
 The reason why we can do this is because the spectrum flattening operation, which is required to derive point $(i)$, preserves the time reversal symmetry. This last Hamiltonian can be mapped into a self-dual Hamiltonian by the replacement of the generators $T\mapsto S$ and, although this changes the invariant $p(\bfk)$, 
 as long as one transforms the time-reversal operator accordingly 
 (this implies changing the matrix $\tau_0\otimes \sigma_y$ appearing in the action of $\mathcal{T}$ in a linear fashion by conjugation by a matrix), the $\mathbb{Z}_2 -$invariant of the relevant vector 
 bundle (associated to time reversal symmetry) is preserved. One can then, because of the form of 
 the self-dual generators and using the relations
  of \eqref{eq: TRS constraints (i)}, describe the system as two time reversal related copies of a $2\times 2$ Hamiltonian. 
 The map $\Phi_2:\bfk\mapsto \vec{h}'(\bfk)/|\vec{h}'(\bfk)|$ has zero winding number because of the condition of time reversal invariance. It follows then, by point $(ii)$ of the Theorem, 
 that the first Chern number will be always identically zero. The $\mathbb{Z}_2 -$invariant for systems which are two copies of time reversal related systems is precisely the first Chern number $\text{mod } 2$, which is, therefore, trivial.
 
The other possibility is to consider that the Hamiltonian (\ref{eq: brunoH}) 
could alternatively be interpreted as being written in the basis
$(\psi_{1\up} \psi_{2\up}  \psi_{1_\dn}\psi_{2\dn})$
\begin{eqnarray}
\mathcal{T}: H(\bfk)\mapsto (\sigma_y\otimes \tau_0)\cdot H^* (-\bfk)\cdot (\sigma_y\otimes \tau_0),
\end{eqnarray}
 where again $\sigma_y\equiv\tau_2$ is used to emphasize the spin  degree of freedom.
Time reversal invariance then implies,
\begin{align}
h(-\bfk)=- h(\bfk),\qquad D(-\bfk)=D(\bfk).
\label{eq: TRS constraints (ii)}
\end{align}
The conditions described above again imply  that $p(\bfk)$ must be negative everywhere 
and also that the winding number of $\Phi_2$ must be zero. As in the preceding case, the resulting model can be seen as to two copies of time reversal related $2\times 2$ Hamiltonians.  The $\mathbb{Z}_2 -$invariant is again trivial. Thus, as claimed, we have proved that the family $H(\bfk)$ can not describe a non-trivial topological insulator.

\subsection{Chern insulator}

The Hamiltonian \eqref{eq: brunoH} may describe  spinless fermions on a lattice
with four orbitals per site, in the basis
$(\psi_1,\psi_2,\psi_3,\psi_4)$. The condition for time-reversal invariance (TRI)
reads $H(\bfk)=H^*(-\bfk)$, which implies:
\begin{eqnarray}
h_{0,1,3}(\bfk) = h_{0,1,3}(-\bfk)\,,&& 
 h_{2}(\bfk) = -h_{2}(-\bfk)\,, \nonumber\\
 D(\bfk)&=&D^*(-\bfk)
 \label{4chern}
\end{eqnarray}
A non-zero Chern number requires violation of any of the conditions (\ref{4chern}).

\subsection{Single band  superconductor}

The Hamiltonian form \eqref{eq: brunoH} can also be read off 
as a Bogoliubov-de Gennes matrix\cite{ludwig} for a single band superconductor
in the particle-hole basis
$(\psi_\up \psi_\dn 
 \psi_\up^\dagger \psi_\dn^\dagger)$. In this case, the matrices $\tau$ in 
 equation \eqref{eq: brunoH} operate in spin space.
 The kinetic energy is $h(\bfk)\cdot\tau\equiv \Xi(\bfk)$.
 Under time-reversal, this kinetic energy transforms
 into the $(2,2)$ block of matrix  \eqref{eq: brunoH},
  which  must then be read off as $-\Xi^T(-\bfk)$.
This  requires that
 $h(\bfk)=h(-\bfk)$.
The off-diagonal term  $id(\bfk) \tau_2\equiv \hat \Delta$ is a 
 spin singlet pairing term. Then, fermionic statistics further
 dictates that $D(\bfk)=D(-\bfk)$.
 Both mappings  $\Phi_{1,2}$ can be nontrivial, 
separated by a topological transition.

If the  Bogoliubov-de Gennes matrix \eqref{eq: brunoH} is interpreted as
written in the Nambu basis 
$( \psi_\up \psi_\dn  \psi_\dn^\dagger -\psi_\up^\dagger)$
then it is of the form
\begin{eqnarray}
\left( \begin{array}{cc}
\hat\Xi & -i\hat \Delta\tau_2\\
i\tau_2\hat\Delta^\dagger & -\tau_2 \hat\Xi^T\tau_2
 \end{array}\right) \,.
 \label{nambu}
\end{eqnarray}
The pairing term 
$
-i\hat\Delta\tau_2 \equiv\psi(\bfk)+\vec d(\bfk)\cdot\vec \tau 
$
where $\psi$ and $\vec d$ denote the amplitudes for singlet and triplet pairing, 
respectively.
By comparing  equations \eqref{eq: brunoH} and (\ref{nambu}) we see that
 \begin{eqnarray}
 h_0(\bfk) = h_0(-\bfk)\,,&&
  \vec h(\bfk) = -\vec h(-\bfk)\,, \nonumber\\
\psi=d_{x,z}=0\,, &&
iD(\bfk)=d_y(\bfk)\,.
 \label{t4}
\end{eqnarray}
Because of the gap condition, the non-trivial mapping can only be $\Phi_2$, in the case where $d_y(\bfk)$ is complex,
of the type $p+ip$ pairing.

\subsection{Multiband  superconductor}

We now assume spin $\up$ electrons to live on a lattice
with two orbitals per site and have kinetic energy
$\Xi_\up(\bfk)=h(\bfk)\cdot\tau$ where the Pauli matrices $\tau$
operate in orbital (or pseudospin) space.
The  spin $\dn$ electrons have kinetic energy
$\Xi_\dn(\bfk)=\Xi_\up^*(-\bfk)=h(-\bfk)\cdot\tau^T$.
We further take the pairing matrix of the form
$\hat\Delta= \left[ \psi(\bfk)+ \vec d(\bfk)\cdot\vec\sigma \right] i\sigma_2i\tau_2$,
where the Pauli matrices $\sigma_i$ operate on physical spin.
 The fermionic
statistics imposes that  $\vec d$ is an even function of $\bfk$, while
$\psi$ is odd.
The Bogoliubov-de Gennes matrix  in the particle-hole basis
$( \psi_{1\up}  \psi_{2\up}  \psi_{1\dn}  \psi_{2\dn} 
\psi_{1\up}^\dagger  \psi_{2\up}^\dagger  \psi_{1\dn}^\dagger  \psi_{2\dn}^\dagger)$
takes the form:
\begin{widetext}
\begin{eqnarray} 
H_4=
\left( \begin{array}{cccc}
h(\bfk)\cdot\tau & 0 & (-d_x+id_y)i\tau_y  &  (\psi+d_z) i\tau_y \\ 
0 & h(-\bfk)\cdot\tau^T  &   (-\psi+d_z)i\tau_y  & (d_x+id_y) i\tau_y  \\
 (d_x^*+id_y^*)i\tau_y & (\psi^*-d_z^*)i\tau_y & - h(-\bfk)\cdot\tau^T  & 0\\
 -(\psi^*+d_z^*) i\tau_y & (-d_x^*+id_y^*) i\tau_y & 0 &-h(\bfk)\cdot\tau   
\end{array}\right) 
\label{ph13}
\end{eqnarray}
\end{widetext}
There are two cases in which the matrix $H_4$ decouples into
two independent blocks:

(I) $\psi=d_z=0$: then the decoupled matrices have the form 
 \eqref{eq: brunoH} if the function $h(\bfk)$ is even,
and $D(\bfk)=-d_x+id_y$ describes spin triplet pairing. 
Since  $\vec d$ is an even function of $\bfk$, this pairing
has odd parity under inversion\cite{Fu1}:
${\cal I}:\hat\Delta(\bfk)\rightarrow  \tau_x \hat\Delta(-\bfk) \tau_x\  =
-\hat\Delta(\bfk)$.
Both mappings  $\Phi_{1,2}$ can be nontrivial and
separated by a topological transition.

(II) 
$d_x=d_y=0$: the general case would be a superposition of singlet ($\psi$)
and triplet ($d_z$) pairings.
The decoupled BdG matrix has the form \eqref{eq: brunoH} if $h_y=0$, which implies
that the mapping $\Phi_1$ for the normal system is trivial. 
A non-trivial  $\Phi_2$ mapping
can be achieved through  spin singlet $p+ip$, 
or triplet $s+id$ pairings,
for instance.

\subsection{Model examples}

In order to illustrate the case (I) above, 
we may write the kinetic energy for $\up$-spin electrons with even $h(\bfk)$,
where $h_0=-t  \left( \cos k_x + \cos k_y \right)$ and 
\begin{eqnarray}
h_x &=&
  t_1  \left( \cos k_x + \cos k_y \right)\,,
\nonumber\\
h_y 
&=&  t_1  \left( \cos k_x - \cos k_y \right)   \,,  \label{model1}\\
h_z &=& t_2 \sin k_x \sin k_y +    \delta
\nonumber
 \,.
\end{eqnarray}
 If  $|t_2| > |\delta |$ then  $C=\pm 2$, 
 otherwise $C=0$.
The term $h_y$  breaks TRS and
is responsible for a non-zero Chern number. The terms  $t_2$ and $h_y$ break
spatial  inversion symmetry.
The amplitude of $h_0$ must be relatively small so as to have a band gap in the normal system.
For instance, the choice $t=0.1$, $t_2=-0.9$, $\delta=0.9$ yields $C=+2$.

As for the function $D(\bfk)=-d_x+id_y$, we consider 
odd parity
spin triplet $d_{x^2-y^2}+i d_{xy}$ pairing:
\begin{eqnarray}
D(\bfk)&=& \Delta_0 + 
\Delta_s (\cos k_x - \cos k_y ) + 
i \Delta_d \sin k_x \sin k_y\nonumber\\
\end{eqnarray}
The choice $\Delta_0=0$, $\Delta_s=0.1$, $\Delta_d =-0.1$, for instance,
 yields  $C=-2$.

\section{Summary and Conclusions}
\label{sec: IV}

We have introduced a Hamiltonian for fermions on a lattice which can describe several physical systems. A theorem regarding this type of Hamiltonians was proved which allows for direct computation of the associated topological invariants.
As a consequence of this theorem, we have shown that this Hamiltonian form cannot describe a non-trivial topological insulator. Nevertheless, it can describe non-trivial Chern insulators, single band superconductors and multiband superconductors. This is an improvement with respect to the usual single band models which are an over simplification because topological materials are, in a realistic physical setup, multiorbital.

One can now imagine a physical system as described,
 and couple it to the same physical system but with different topological invariants. 
 By the bulk-boundary correspondence principle, there will exist edge-states at the boundary which, in the superconducting cases, will be Majorana fermions. Physically, this model potentially describes heterojunctions of topological-insulator/superconductor (TI/Sc heterojunctions). 
 It can be regarded as  an extension of previous suggestions  for practical realization of Majorana modes in 
  TI/Sc heterojunctions\cite{alicea} which did not identify the physical significance of the topological criterion found 
 as a $\mathbb{Z}_2-$ topological invariant $p(\bfk)$ (the Pfaffian polynomial). 
 The topological invariant describing the system is not only the first Chern number, but also the sign of the Pfaffian polynomial coming from this notion of duality we have discussed.  If this behaviour is realized
 experimentally, then this would allow for potential applications in topological quantum computation as this could be a way to control Majorana fermions.

\section*{Acknowledgments}

We acknowledge financial support from 
Funda\c{c}\~ao para a Ci\^encia e Tecnologia 
(Project EXPL/FIS-NAN/1728/2013 and Grant No. UID/CTM/04540/2013). In addition, B.M. acknowledges the support from Funda\c{c}\~{a}o para a Ci\^{e}ncia e a Tecnologia (Portugal), namely through programmes PTDC/POPH and projects UID/Multi/00491/2013, UID/EEA/50008/2013, IT/QuSim and CRUP-CPU/CQVibes, partially funded by EU FEDER, and from the EU FP7 projects LANDAUER (GA 318287) and PAPETS (GA 323901). B. M. thanks M. Abreu for very fruitful discussions regarding the mathematical concepts behind the paper.


\end{document}